\documentclass[12pt]{iopart}

\begin{document}
\maketitle
\title[Complete Integrability Of The Fifth-Order MNW system]{Complete Integrability Of The Fifth-Order Mikhailov-Novikov-Wang system}

\author[Daryoush TALATI]{Daryoush TALATI}

\ead{talati@eng.ankara.edu.tr , daryoush.talati@gmail.com}

\begin{abstract}

The symplectic-Hamiltonian formulation and recursion operator of the fifth-order Mikhailov-Novikov-Wang system are given.

\end{abstract}

The Hamiltonian and symplectic structures are well known to be of paramount importance, as the whole conserved quantity hierarchy and symmetry hierarchy for the system in question are then readily generated by the repeated application
of these operators to the seed symmetry. Complete integrability of an equation is established by
proving a compatible bi-Hamiltonian, (or bi-symplectic or Hamiltonian-symplectic operators) defining the Magri scheme of infinite symmetry hierarchy $u_{t_{i}}=F_{i}[u]$ all of which is in conservation law form
\begin{eqnarray*}\fl
\begin{array}{l}
u_{t_{i}}=F_{i}[u]={\rm K}G_{i}[u]={\rm J}G_{i+1}[u], \;\; G_{i}=\mathrm{\delta}(\rho_{i}[u]),\;\;\;i=-1,0,1,2,3,\cdots
\end{array}
\end{eqnarray*}
with respect to two compatible Hamiltonian (skew-adjoint, Jacobi identity satisfying) operators $\mathrm{J}$ and $\mathrm{K}$ by the conserved density $\rho_i[u]$ which are in involution.

Recently, Mikhailov, Novikov, and Wang \cite{a} applied the symmetry analysis to a class of fifth-order nonlinear equations, and found two new cases that pass the symmetry test:

\begin{eqnarray} \fl 
\left\{\begin{array}{ll} 
p_t=&-\frac{5}{3}p_5 -10qq_3-15q_1q_2+10pp_3+25p_1p_2-6q^2q_1+6q^2p1\\& +12pqq_1-12p^2p_1,\\
q_t=&15q_5+30q_1q_2-30q_3p-45q_2p_1-35q_1p_2-10qp_3-6q^2q_1+6q^2p_1\\&+12p^2q_1+12qpp_1.
\end{array}\right. \label{k}
\end{eqnarray}

 \begin{eqnarray} \fl 
\left\{\begin{array}{ll} 
p_t=&(9-5\sqrt{3})p_{5}+D_x\{2(9-5\sqrt{3})pp_{2}+(-12+7\sqrt{3})p_{1}^2 \}+2(3-\sqrt{3})p_{4}q\\&+2(6-\sqrt{3})p_{2}q_{1}+2(3-2\sqrt{3})p_{1}q_{2}-6(1+\sqrt{3})pq_{3}+D_x\{2(33+19\sqrt{3})qq_{2}\\&+(21+12\sqrt{3})q_{1}^2 \}+\frac{4}{5}(-12+7\sqrt{3})p^2p_{1}+\frac{8}{5}(3-2\sqrt{3})(qpp_{1}+p^2q_{1})\\&+\frac{4}{5}(24+13\sqrt{3})q^2p_{1}+\frac{8}{5}(36+20\sqrt{3})pqq_{1}-\frac{8}{5}(45+26\sqrt{3})q^2q_{1},\\ 
q_t=&(9+5\sqrt{3})q_{5}+D_x\{2(33-19\sqrt{3})pp_{2}+(21-12\sqrt{3})p_{1}^2 \}-6(1-\sqrt{3})p_{4}q\\&+2(3+2\sqrt{3})p_{2}q_{1}+2(6+\sqrt{3})p_{1}q_{2}+2(3+\sqrt{3})pq_{3}+D_x\{2(9+5\sqrt{3})qq_{2}\\&-(12+7\sqrt{3})q_{1}^2 \}-\frac{8}{5}(45-26\sqrt{3})p^2p_{1}+\frac{8}{5}(36-20\sqrt{3})qpp_{1}\\&+\frac{4}{5}(24-13\sqrt{3})p^2q_{1} +\frac{8}{5}(3+2\sqrt{3})(q^2p_{1} +pqq_{1})-\frac{4}{5}(12+7\sqrt{3})q^2q_{1}.
\label{man}
\end{array}\right.
\end{eqnarray}

Here $p_i=\frac{\partial^ip}{\partial x^i},~q_i=\frac{\partial^iq}{\partial x^i}$. System (\ref{k}) admits a reduction $q=0$ to the Kaup-Kupershmidt equation $p_t=p_{5x}+10pp_{3x}+25p_{x}u_{xx}+20p^2p_{x}$. 
Bi-Hamiltonian structure for system (\ref{k}) was discussed in \cite{v}. However, no recursion operator, symplectic or Hamiltonian structure for (\ref{man}) was known. It is important to explore properties of system(\ref{man}) in order to find out whether it enjoys any features such as Hamiltonian structure and recursion operator too. Working with the present form of the system (\ref{man}) is an obstinate problem. In the variables $ p= \frac{3+\sqrt{3}}{2}u - \sqrt{3}v,~~q= \frac{3-\sqrt{3}}{2} u + \sqrt{3}v$ system (\ref{man}) is converted to

\begin{eqnarray} \fl 
\left\{\begin{array}{ll} 
u_t=&u_{5}+\frac{5}{2}v_{5}+6u_{3}u+18u_{3}v+12v_{3}u+42v_{3}v+12u_{2}u_{1}+24u_{2}v_{1}+21v_{2}u_{1}\\&+42v_{2}v_{1}+\frac{54}{5}u_{1}u^2+\frac{108}{5}u_{1}uv-18u_{1}v^2+\frac{72}{5}v_{1}u^2-\frac{72}{5}v_{1}uv-144v_{1}v^2 ,
\\\\
v_t=&\frac{5}{4}u_{5}+\frac{7}{2}v_{5}+3u_{3}u+6v_{3}u-6v_{3}v+\frac{3}{2}u_{2}u_{1}-6u_{2}v_{1}-3v_{2}u_{1}-33v_{2}v_{1}\\&+\frac{36}{5}u_{1}v^2-\frac{18}{5}v_{1}u^2-\frac{36}{5}v_{1}uv+\frac{126}{5}v_{1}v^2 .
\end{array}\right.
\end{eqnarray}

Here and below $u_i=\frac{\partial^iu}{\partial x^i},~v_i=\frac{\partial^iv}{\partial x^i}$. We prefer this form, since all its coefficients are integer-valued. This system possesses generalized symmetries of orders 7, 11, 13 and 17, as well as
conserved densities of weights 2, 4, 8, 10, 14 and 16. We list some of the simplest associated conserved densities,

\begin{eqnarray*} \fl 
\begin{array}{ll}

\mathcal{\rho}_{1} &=u+v, \\
\rho_{2}& =-15u_{2}u -80v_{2}u -110v_{2}v -8u^{3} -24u^{2}v +24uv^{2} +88v^{3} ,\\
\rho_{3} &=-100u_{4}u -550v_{4}u -750v_{4}v - 125u_{2}u^{2} -250u_{2}uv +130u_{2}v^{2} -335v_{2}u^{2} \\& -580v_{2}uv +550v_{2}v^{2} -40u^{4} -160u^{3}v +48u^{2}v^{2} +704uv^{3} +536v^{4} ,\\

\rho_{4} &= 102375u_{4}u^{3}+57750u_{6}u^{2}+261675u_{2}^{2}u^{2}-3309120v_{2}uv^{3}-1730160u_{2}v^{4}\\&-427680v_{1}^{2}u^{3}+20736u^{6}+124416u^{5}v+103680u^{4}v^{2}-622080u^{3}v^{3}\\&-1181952u^{2}v^{4}+165888uv^{5}+1000(390u_{8}v+390v_{8}u+1065v_{8}v\\&+324u_{6}uv+1467v_{4}u_{2}u+3294v_{4}v_{2}u+2319v_{4}v_{2}v+876u_{3}v_{1}u^{2}-327v_{3}v_{1}u^{2}\\&+7008v_{3}v_{1}uv+1314u_{2}v_{2}u^{2}+2880v_{2}^{2}uv-1134v_{2}u^{2}v^{2})+100(1425u_{8}u\\&-1962u_{6}v^{2}-207v_{6}u^{2}+144v_{6}uv-165v_{6}v^{2}+3096u_{5}v_{1}u+324v_{5}v_{1}u\\&+2085u_{4}u_{2}u+5688u_{4}u^{2}v+3744u_{4}uv^{2}-7216u_{4}v^{3} +6567v_{4}u^{2}v\\&+12504v_{4}uv^{2}-23715v_{4}v^{3}+21972u_{3}v_{1}uv+2812v_{4}u^{3}+21078u_{2}v_{2}uv\\&+1458u_{2}u^{4}+5832u_{2}u^{3}v+2916u_{2}u^{2}v^{2}-12312u_{2}uv^{3}+1035v_{2}^{2}u^{2}\\&+36225v_{2}^{2}v^{2}+2646v_{2}u^{4}+6696v_{2}u^{3}v-32508v_{2}v^{4}-21384v_{1}^{2}u^{2}v+10998v^{6}) ,\\
\dot{.}&\\
\dot{.}&\\
\end{array}
\end{eqnarray*}

Therefore the system (\ref{man}) is very likely to be integrable. So we are now in the position to construct a symplectic-Hamiltonian structure\cite{mag,10}

\emph{{\bf Theorem 1:}} The infinite hierarchy of System (\ref{man}) have infinite conserved densities together suffice to write Magri schemes with the compatible pair of Hamiltonian operator 

\begin{eqnarray} 
\mathrm{K} = \left(
\begin{array}{cc} 
\mathrm{K} _{1}&\mathrm{K} _{2}\\
-\mathrm{K _{2}}^{*}&\mathrm{K} _{4}
\end{array}\right) 
\end{eqnarray}

where
\begin{eqnarray*} \fl 
\begin{array}{ll}
K_1 =&-60D_x^{3}+72(v-u)D_x+72D_x(v-u)\\
K_2 =&15D_x^{3}+36(u-4v)D_x+18( u_{1}-4v_{1})\\
K_4 =&-15D_x^{3}+18(v-u)D_x+18D_x(v-u) 
\end{array}
\end{eqnarray*}

and symplectic operator

\begin{eqnarray}
\mathrm{J} = \left(
\begin{array}{cc} 
\mathrm{J} _{1}&\mathrm{J} _{2}\\
-\mathrm{J _{2}}^{*}&\mathrm{J} _{4}
\end{array}\right)
\end{eqnarray}

where

\begin{eqnarray*} \fl 
\begin{array}{ll}
J_1 =& \alpha_{0} D_x^9 + \alpha_{1} D_x^7 + D_x^7 \alpha_{1} + \alpha_{2} D_x^5 + D_x^5 \alpha_{2} + \alpha_{3} D_x^3 + D_x^3 \alpha_{3} + \alpha_{4} D_x\\& + D_x \alpha_{4} + \beta_1 D_x^{-1} \beta_7 + \beta_7 D_x^{-1} \beta_1 + \beta_3 D_x^{-1} \beta_5 + \beta_5 D_x^{-1} \beta_3 \\

J_2=& \alpha_{5} D_x^9 + D_x^7 \alpha_{6} + \alpha_{7} D_x^6 + D_x^5 \alpha_{8} + D_x^4 \alpha_{9} 
 + D_x^3 \alpha_{10} + D_x^2 \alpha_{11} + D_x \alpha_{12}\\& + \alpha_{13}+ \beta_1 D_x^{-1} \beta_8 + \beta_7 D_x^{-1} \beta_2 + \beta_3 D_x^{-1} \beta_6 + \beta_5 D_x^{-1} \beta_4 \\

J_4 =& \alpha_{14} D_x^9 + \alpha_{15} D_x^7 + D_x^7 \alpha_{15} + \alpha_{16} D_x^5 + D_x^5 \alpha_{16} + \alpha_{17} D_x^3 + D_x^3 \alpha_{17} + \alpha_{18} D_x\\& + D_x \alpha_{18} + \beta_2 D_x^{-1} \beta_8 + \beta_8 D_x^{-1} \beta_2 + \beta_4 D_x^{-1} \beta_6 + \beta_6 D_x^{-1} \beta_4 .
\end{array}
\end{eqnarray*}

By straighthforward but tedious computation it can be readily verified that the functional trivector of linear combination $KJK+\lambda K$ with constant $\lambda$, vanishes independently from the value of $\lambda$ \cite{OLV}. The explicit form of $\alpha_i$ and $\beta_i$ are :

$\fl\\
\beta_1 =\frac{1}{6}(\delta_{u}\int\rho_1dx),~
\beta_2 =\frac{1}{6}(\delta_{v}\int\rho_1dx),~
\beta_3 =9(\delta_{u}\int\rho_2dx),~
\beta_4 =9(\delta_{v}\int\rho_2dx)\\
\beta_5 =(\delta_{u}\int\rho_3dx),~
\beta_6 =(\delta_{v}\int\rho_3dx),~
\beta_7 =(\delta_{u}\int\rho_4dx),~
\beta_8 =(\delta_{v}\int\rho_4dx)\\
\alpha_{0}=46875, \alpha_{5}=128125, \alpha_{14}=350000\\
\alpha_{1}=118750u +118750v, \alpha_{6}=537500u +395000v, \alpha_{15}=580000u +190000v\\
\alpha_{2}= 1000(-340u_{2} -415v_{2} +237u^{2} +474uv +174v^{2})\\
 \alpha_{3}=886250u_{4}+1358750v_{4}-893250u_{1}^{2}+100(-2970u_{2}u-2970u_{2}v-4050v_{2}u-8640v_{2}v-20115u_{1}v_{1}-18000v_{1}^{2}+2760u^{3}+8280u^{2}v+3744uv^{2}-4368v^{3}\\
\alpha_{4}=100(-4750u_{6}-6550v_{6}+5130u_{4}u+5130u_{4}v+9570v_{4}u+10380v_{4}v+17340u_{3}u_{1}+18825u_{3}v_{1}+19950v_{3}u_{1}+27780v_{3}v_{1}+16695u_{2}^{2}+37530u_{2}v_{2}+2328u_{2}u^{2}+4656u_{2}uv+816u_{2}v^{2}+26910v_{2}^{2}+4200v_{2}u^{2}+3504v_{2}uv-4800v_{2}v^{2}-7038u_{1}^{2}u-7038u_{1}^{2}v-15588u_{1}v_{1}u-9216u_{1}v_{1}v-15768v_{1}^{2}u-9720v_{1}^{2}v)+184320u^{4}+737280u^{3}v+449280u^{2}v^{2}-956160uv^{3}-696960v^{4}\\
\alpha_{7}= -2263750u_{1} -2230000v_{1}\\
\alpha_{8}= -8701250u_{2} +100(-90650v_{2} +8100u^{2} +9720uv -1800v^{2})\\
\alpha_{9}=27221250u_{3} +100(277200v_{3} -53760u_{1}u -39540u_{1}v -53400v_{1}u -22080v_{1}v)\\
\alpha_{10}= 100(-384075u_{4} -373050v_{4} +94800u_{2}u +74460u_{2}v +87960v_{2}u +14160v_{2}v +79845u_{1}^{2} +145200u_{1}v_{1} -4710v_{1}^{2} +7824u^{3} +12960u^{2}v -22176uv^{2} -40704v^{3})\\
\alpha_{11}=29121250u_{5} +100(267400v_{5} -90630u_{3}u -69660u_{3}v -81360v_{3}u +14760v_{3}v -216345u_{2}u_{1} -194700u_{2}v_{1} -181200v_{2}u_{1} +97620v_{2}v_{1} -46584u_{1}u^{2} -63072u_{1}uv +20448u_{1}v^{2} -48096v_{1}u^{2} -6048v_{1}uv +124992v_{1}v^{2})\\
\alpha_{12}=100(-113250u_{6}-93375v_{6}+55830u_{4}u+41880u_{4}v+62100v_{4}u-9000v_{4}v+156345u_{3}u_{1}+143430u_{3}v_{1}+132270v_{3}u_{1}-99570v_{3}v_{1}+110295u_{2}^{2}+190260u_{2}v_{2}+53256u_{2}u^{2}+73824u_{2}uv-24576u_{2}v^{2}-84600v_{2}^{2}+59808v_{2}u^{2}+8448v_{2}uv-149856v_{2}v^{2}+80100u_{1}^{2}u+60120u_{1}^{2}v+154800u_{1}v_{1}u+10080u_{1}v_{1}v-11232v_{1}^{2}u-230472v_{1}^{2}v)+437760u^{4}+852480u^{3}v-2833920u^{2}v^{2}-6612480uv^{3}-1601280v^{4}\\
\alpha_{13}=100(17025u_{7}+10050v_{7}-16860u_{5}u-11820u_{5}v-23820v_{5}u+1740v_{5}v-56655u_{4}u_{1}-53460u_{4}v_{1}-61500v_{4}u_{1}+21750v_{4}v_{1}-94005u_{3}u_{2}-82530u_{3}v_{2}-21480u_{3}u^{2}-28992u_{3}uv+8904u_{3}v^{2}-83520v_{3}u_{2}+92520v_{3}v_{2}-25296v_{3}u^{2}-624v_{3}uv+56640v_{3}v^{2}-98292u_{2}u_{1}u-74424u_{2}u_{1}v-94728u_{2}v_{1}u-6168u_{2}v_{1}v-100200v_{2}u_{1}u-5016v_{2}u_{1}v+13416v_{2}v_{1}u+274056v_{2}v_{1}v-27558u_{1}^{3}-75816u_{1}^{2}v_{1}+16668u_{1}v_{1}^{2}+112392v_{1}^{3}+62208v_{1}uv^{2})-1255680u_{1}u^{3}-2453760u_{1}u^{2}v+2142720u_{1}uv^{2}+3928320u_{1}v^{3}-1359360v_{1}u^{3}-622080v_{1}u^{2}v+3271680v_{1}v^{3}\\
\alpha_{16}=100 (-26725u_{2} -33775v_{2} +5640u^{2} -3120uv -11280v^{2}\\
\alpha_{17}=100(62525u_{4} +89150v_{4} -10740u_{2}u +13380u_{2}v -16860v_{2}u -8220v_{2}v -21240u_{1}^{2} -16920u_{1}v_{1} -62190v_{1}^{2} +5280u^{3} -288u^{2}v -41472uv^{2} -53184v^{3})\\
\alpha_{18}= 100(-36400u_{6} -50050v_{6} +15000u_{4}u -2460u_{4}v +26820v_{4}u +10260v_{4}v +53190u_{3}u_{1} +15300u_{3}v_{1} +54540v_{3}u_{1} +101880v_{3}v_{1} +45060u_{2}^{2} +60600u_{2}v_{2} +3696u_{2}u^{2} +480u_{2}uv -19200u_{2}v^{2} +113550v_{2}^{2} +6864v_{2}u^{2} -11904v_{2}uv -47280v_{2}v^{2} -9216u_{1}^{2}u +1008u_{1}^{2}v -2592u_{1}v_{1}u +26064u_{1}v_{1}v -39096v_{1}^{2}u -30600v_{1}^{2}v)+253440u^{4} -92160u^{3}v -3179520u^{2}v^{2} -2027520uv^{3} +4746240v^{4}\\
$

One of the interesting points here is that System (\ref{man}) does not admit any reduction to any of KdV, Sawada-Kotera or Kaup-Kupershmidt equations. Recursion operator for this system could be factorized as $R=K^{-1}J$ where $K^{-1}$ is the inverse Hamiltonian operator.

\end{document}